\documentclass[fleqn,usenatbib,useAMS]{mnras}
\usepackage[T1]{fontenc}
\usepackage{ae,aecompl}
\usepackage[running]{lineno}
\usepackage{graphicx}
\usepackage{amsmath}
\usepackage{amssymb}
\usepackage{hyperref}
\usepackage{multirow}
\usepackage{txfonts}
\usepackage{calligra}
\usepackage{xcolor}
\usepackage{physics}

\newcommand{\black}{\color{black}}

\title[Bayesian evidence and dark energy]{Bayesian evidence and model selection approach for time-dependent dark energy}

\author[M. Khorasani et al.]{
Mohsen Khorasani\thanks{Contact e-mail: \href{mkhorasani@shirazu.ac.ir}{mkhorasani@shirazu.ac.ir}}$^{1,2}$,
Moein Mosleh\thanks{Corresponding author e-mail: \href{moein.mosleh@shirazu.ac.ir}{moein.mosleh@shirazu.ac.ir}}$^{1,2}$,
Ahmad Sheykhi$^{1,2}$
\\
$^{1}$Department of Physics, School of Science, Shiraz University, Shiraz 71946-84795, Iran\\
$^{2}$Biruni Observatory, School of Science, Shiraz University, Shiraz 71946-84795, Iran}
\date{}
\begin{document}
\pagerange{\pageref{firstpage}--\pageref{lastpage}}
\pubyear{2021}
\maketitle
\label{firstpage}
\setlength{\parindent}{10pt}
\begin{abstract}
\black
We use parameterized post-Friedmann (PPF) description for dark energy and apply ellipsoidal nested sampling  to perform the Bayesian model selection method on different time-dependent dark energy models using a combination of  $Planck$ and data based on distance measurements, namely baryon acoustic oscillations and supernovae luminosity distance. Models with two and three free parameters described in terms of linear scale factor $a$, or scaled in units of e-folding $\ln a$ are considered.
Our results show that  parameterizing dark energy in terms of $\ln a$ provides better constraints on the free parameters than polynomial  expressions. In general,  two free-parameter models are adequate to describe  the dynamics of the dark energy compared to their three free-parameter generalizations. According to the Bayesian evidence,  determining the strength of support for  cosmological constant $\Lambda$  over polynomial dark energy models  remains inconclusive. Furthermore, considering the $R$ statistic as the tension metric shows that one of the polynomial models gives rise to a tension between $Planck$ and distance measurements data sets.  The preference for the logarithmic equation of state over $\Lambda$ is inconclusive, and the strength of support  for $\rm \Lambda$CDM over the oscillating model is moderate.
\black
\end{abstract}
\black
\section{Introduction}
Observational probes based on measuring distance, namely, Type Ia supernovae as a standard candle, directly imply that the Universe is experiencing an accelerating phase. In the standard model of cosmology, explaining this acceleration translates into the requirement that the dominant budget of the Universe contains a strange form of stress-energy, dubbed dark energy  \citep{perlmutter1998discovery,riess1998observational}.

There are some candidates for dark energy. The most straightforward choice is the cosmological constant $\Lambda$ introduced by Einstein to establish a static universe \citep{1917SPAW.......142E}. The cosmological constant can be considered as vacuum energy with a constant equation of state $p_{\Lambda} = - \rho_{\Lambda}$. Even though it is just characterized by a single parameter, $\Lambda$CDM cosmology is in excellent agreement with observational data and  still represents a good fit  against a wide range of cosmological probes \citep{Planck:2015bue}. Despite this success, the cosmological constant  suffers from two  problems that confirm $\Lambda$ must have been fine-tuned. Energy density $\rho_{\Lambda}$  is about 120 orders of magnitude smaller than its value  at the Planck  scale $k_{\rm max} = M_{\rm Pl} \approx 10^{19} \rm GeV$  and is still about 40 order of magnitude smaller than the value expected by a cut-off at the Quantum chromodynamics (QCD) scale $k_{\rm max} \approx 0.1 \rm GeV$ \citep{Weinberg:1988cp}.  Another problem arises from the fact that dark energy must be subdominant to let matter perturbations grow and just begin to be dominated after radiation and matter era at the redshift $z\sim1$,  which refers to the coincidence problem.  For two conditions to be met, energy density $\rho_{\Lambda}$ must be fine-tuned.

In order to alleviate problems related to the $\Lambda$ dark energy, time-dependent dark energy has been proposed \citep{Wetterich:1987fm, Ratra:1987rm, Caldwell:1997ii}.  It is worth mentioning that dynamical dark energy does not solve cosmological constant problems, because in this approach it is supposed that $\Lambda$ is zero due to some unknown mechanism, and the evolution of dark energy arises from dynamics of a minimally coupled  scalar field.
Time-varying dark energy is described by the equation of state $\bar p(a)=w(a)\bar \rho(a)$ with fluid sound speed defined in the rest frame of fluid in terms of perturbed pressure and density $\delta p(k,a) = c^2_{s}(k,a) \delta \rho(k,a)$, where bar denotes average background quantity and $k$ is the wave number.

In a perturbed and inhomogeneous universe, the scalar part of the off-diagonal space-space stress-energy tensor perturbation is described by anisotropic stress $\Pi(k,a)$.
Assuming zero curvature for simplicity, the geometry of space-time is described by Hubble expansion rate $H$ and  two components of metric perturbations $\Phi$ and $\Psi$.
Thus, cosmological probes impose constraints on one {\it background} parameter $H(a)$ and two {\it perturbations} function of scale and time $\Phi(k,a)$ and $\Psi(k,a)$ \citep{Kunz:2012aw}.  Dynamical dark energy which gives rise to different expansion rate $H(a)$ and metric perturbations $\Phi(k,a)$ and $\Psi(k,a)$,  provides different growth of structure and cosmic microwave background (CMB) anisotropies on cosmological scales with respect to $\Lambda$CDM cosmology. The functions $\{H,\Phi,\Psi\}$ exploring metric and geometry of space-time are equivalent to the functions $\{w,c^2_{s},\Pi\}$ describing  physical characteristic of time-dependent dark energy \citep{Ballesteros:2011cm}.

Nevertheless, time-dependent dark energy suffers from some problems. Dark energy with equation of state $w<-1$ called  phantom dark energy model \citep{caldwell2002phantom}, and minimally coupled scalar field dark energy model causes  gravitational instabilities whenever it crosses the phantom divide line $w=-1$ \citep{2005PhRvD..71b3515V}. In fact, minimally coupled scalar filed should admit an internal degree of freedom to evade  instabilities. Assuming dark energy fluctuations  are internally adiabatic, perturbations become unstable  when  adiabatic pressure response to density fluctuations
\begin{equation}\label{adiabatic}
\delta p \equiv c^2_{ad}\delta \rho=\left(\frac{{\bar p}^{\prime}}{{\bar \rho}^{\prime}}\right) \delta \rho
=\left(w -\frac{1}{3} \frac{{\rm d}\ln (1+w)}{{\rm d}\ln a}\right)\delta \rho,
\end{equation}
becomes singular at $w=-1$ or negative ($w<0$). This corresponds to an undefined  or imaginary  adiabatic sound speed, where the latter leads to instabilities in the dark energy \citep{Hu:2004kh}. Here, prime denotes derivative respect to conformal time.

To establish a consistent time-varying dark energy while passing the phantom divide line, the PPF description is proposed \citep{Fang:2008sn}.
In the PPF formalism, the density and momentum components of dark energy are replaced by a dynamical variable $\Gamma$ which preserves the conservation of energy-momentum through its equation of motion. Since fluid adiabatic sound  speed $c_{ad}$ or the relationship between pressure and density perturbation of dark energy leads to  instabilities in dark energy,
the PPF formalism replaces this condition on the pressure perturbations with a connection between the momentum density of the dark energy and matter part. This relation is decomposed to super-horizon scales and transition scales under which the dark energy becomes smooth relative to the matter.

There have been proposed various  parameterizations of the equation of state for dynamical dark energy models in the literature. Some models are described in terms of redshift $z$. For example, the model $w(z)=w_0+w_a z$ with condition $w(z) \ge-1$ is the straightforward choice, which provides a good fit for  supernovae measurements not for CMB data \citep{2001PhRvL..86....6M,2004ApJ...607..665R}.  The other model is parameterization $w(z) = w_0 + w_a z/{(1+z)}^p$, assuming $p=1,2$ with the possibility of crossing phantom divide line, which is constrained in light of supernovae data in combination with Wilkinson Microwave Anisotropy Probe (WMAP) CMB anisotropies measurements \citep{2005MNRAS.356L..11J}.
One can express the equation of state in terms of the scale factor $a$. The famous model is $w(a)=w_0+w_a(1-a)$ where is constrained in favour of $Planck$ CMB anisotropies measurements in combination with weak lensing  and non-CMB data \citep{Planck:2015bue,Planck:2018vyg}.
It is also customary to describe the evolution of dark energy in units of $\ln a$, which includes two classes of logarithmic and oscillating models.  The model $w(a)=w_0+w_a\sin(\mathcal{A}\ln a + \theta)$ is constrained by WMAP data in combination with supernovae luminosity measurements \citep{2006PhRvD..74h3521X}. Some different models of oscillating models with two free parameters are considered for $Planck$ in combination with data based on distance measurement \citep{2018PhRvD..98f3510P}. The logarithmic model $w(a) = w_0 + w_a \ln a$ is constrained in light of data based on distance  measurements such as supernovae and baryon acoustic oscillations alongside Hubble parameter  measurements \citep{2022A&A...668A.135S}.

The above models have been constrained by different data sets which makes the comparison of models with each other impossible. Furthermore, a few of them have been described in the PPF formalism. There will occur inconsistencies between the Einstein equations because of the Bianchi identities if one artificially turns the dark energy perturbations off while crossing phantom divide line \citep{Fang:2008sn}.
Therefore, we apply PPF formalism alongside unique data sets to compare different time-dependent dark energy models with each other. 
The Bayes factor $\ln B$ plays a crucial role in the comparison model approach \citep{jeffreys1998theory}. Hence,  we perform an ellipsoidal nested sampling algorithm which allows us to infer the Bayes factor alongside estimating the free parameters.
We follow the conventional approach and choose the minimally-coupled scalar field, called quintessence to describe the dynamics of dark energy. For quintessence dark energy, the rest frame sound speed is set to $c^2_s =1$, and anisotropy stress $\Pi(k,a)$ vanishes. Due to relativistic sound speed, inside the horizon, dark energy density perturbations are suppressed and are comparable to matter perturbations only on the super-horizon scales.

The paper is organized as follows. In section \ref{method-data}, we describe cosmological code and data alongside the sampling method used for estimating parameters and performing the model selection approach.
We impose observational constraints on the time-dependent dark energy models in section \ref{result} and summarize our results based on the dispersion of posterior distributions and Bayesian model selection method in section \ref{discuss}.

\section{Methodology and Data}\label{method-data}
\subsection{Cosmological codes and inference}
Theoretical predictions of CMB anisotropies  and other cosmological observables are calculated using  Boltzmann codes
$\textsc{Camb}$ \citep{Lewis:1999bs} or $\textsc{Class}$ \citep{Blas:2011rf}. For estimating cosmological parameters, we use public cosmological code ${\textsc{CosmoSIS}}$ \citep{Zuntz:2015dhn} which acts as an interface between {\textsc{CosmoMC}} \citep{Lewis:2002ah} and {\textsc{MontePython}} \citep{Audren:2012wb} based on {\textsc{Camb}} and {\textsc{Class}}, respectively.  With the help of ellipsoidal nested sampling {\textsc{MultiNest}} \citep{Feroz:2008xx,Feroz:2013hea} implemented in {\textsc{CosmoSIS}}, we apply Bayesian inference not only to infer the posterior probability distributions of the model parameters but also perform model selection approach.

In contrast to ordinary Markov chain Monte Carlo (MCMC) sampling, namely Metropolis-Hastings, ellipsoidal nested sampling {\textsc{MultiNest}} can simultaneously estimate the Bayesian evidence $\mathcal{Z}_i$ of a model $\mathcal{M}_i$, allowing the comparison between different models via the Bayes factor  $B_{12}=\mathcal{Z}_1/\mathcal{Z}_2$ \citep{jeffreys1998theory}.
The Bayes factor with value $B_{12}>1$ or $B_{12}<1$ implies that model with evidence $\mathcal{Z}_1$ or  $\mathcal{Z}_2$ has the highest evidence, respectively. According to Jeffreys's scale, if $|\ln B_{12}| < 1$ there is no significant preference for the model with the highest evidence, if $1<|\ln B_{12}|<2.5$ the preference for the highest evidence is weak to moderate, if $2.5 < |\ln B_{12}| < 5$ the preference is moderate to strong, and if $|\ln B_{12}| > 5$ it is strong \citep{2007MNRAS.378...72T}.

\subsection{Data}
In this subsection, we discuss the data sets we use, from $Planck$ CMB angular anisotropies measurements alongside non-CMB data.
The CMB provides the cleanest probe of large scales to investigate characteristics of dark energy. However, at the largest scales constraints coming from $Planck$ CMB power spectra are weak due to cosmic variance. Meanwhile, non-CMB data sets are useful for imposing constraints on late-time cosmology. The combination of $Planck$ CMB  angular anisotropies power spectra  and non-CMB data allows breaking degeneracies from concordance $\Lambda$CDM cosmology at the low redshifts.
\subsubsection{CMB data}

We are concerned about the scalar perturbations. The dominant part in the CMB temperature power spectrum comes from lensing and the integrated Sachs-Wolfe (ISW) effect. $Planck$ CMB angular anisotropies measurements  \citep{Planck:2019nip} based on the Planck Release 3 or 2018 likelihoods include temperature, polarization, and CMB lensing  power spectra. We apply polarization data alongside the temperature power spectrum to form baseline low-$\ell$ likelihood. 
We neglect CMB lensing because the inclusion of CMB lensing does not have a significant impact on the constraints of the parameters \citep{Planck:2015bue, Planck:2018vyg}. Temperature data (TT) includes  low-$\ell$ likelihood {\tt Commander} with multipoles in range $2 \le \ell \le 29$, and high-$\ell$ likelihood $\tt Plik_{-}lite$ which covers multipoles in range $30 \le \ell \le 2508$. Polarization data (EE) consists of {\tt SimAll} likelihood $\tt lowE$ which spans multipole from $\ell=2$ to $\ell=29$. We use the notation $Planck$ TT+lowE for the combination of three likelihoods.

To decrease run time,  we use marginalized high-$\ell$ $\tt Plik_{-}lite$ likelihood not the high-$\ell$ likelihood $\tt Plik$ with 16 nuisance parameters, because the former is similar to  both low-$\ell$ likelihoods consists of  just one nuisance parameter $\tt A_{-}Planck$.

\subsubsection{Non-CMB data}
Non-CMB measurements consist of two different probes, perturbation and background data sets. Perturbation data sets which are measuring the evolution of the density perturbations provide an independent test to investigate the nature of dark energy
because it changes the expansion rate of the Universe and hence the growth rate of structures in it \citep{2005PhRvD..72d3529L}.
Late time-probes, namely, cosmic shear, galaxy clustering, and galaxy-galaxy lensing are suited for determining present-day matter density $\Omega_m$ and parameter $\sigma_8$ \citep{2018PhRvD..98d3526A}. These probes
are based on the two-point correlation function which results in being used as an indirect clue to explore the nature of the dark energy. The quantity $\sigma_8$ is  the amplitude of the power spectrum of matter fluctuations averaged on the sphere of radius $R=8h^{-1}$Mpc, where $h$ denotes the Hubble constant in the unit of 100 $\rm km~s^{-1}Mpc^{-1}$.
Since the inclusion of  probes based on a two-point function due to suppression of the perturbations of smooth dark energy on small scales provides poor constraints on the dark energy parameters \citep{Planck:2015bue}, we do not apply these data sets in our analysis. 

Background data sets based on distance measurements, namely, Type Ia supernovae and baryon acoustics oscillations provide a direct clue for exploring the expansion history of dark energy. We only use these data sets alongside redshift space distortion of galaxies as non-CMB data sets for our analysis. These probes are called geometrical data\footnote{Redshift space distortions measurements belong to perturbation data sets whose evolution equations for density perturbations are second-order in time. Nevertheless, we consider this data set alongside data based on distance measurements as geometrical data.} because they measure the large-scale geometry of space-time, and their interpretation relies on energy conservation and the first-time derivative of the expansion scale factor \citep{bertschinger2006growth}.

\subsubsection*{Baryon acoustic oscillations}
Baryon acoustic oscillations (BAO) which represent oscillations in the baryon-photon plasma in advance of  the recombination era on the matter power spectrum, lead to the acoustic peak structure of the CMB power spectra. These oscillations can be calibrated to the sound horizon at the end of the drag epoch and remain imprinted into the matter distribution up to now \citep{Eisenstein:1998tu}.

The sound horizon scale measured by BAOs at 147 Mpc, makes BAO measurements insensitive to nonlinear physics.
This feature implies that  BAOs as primary non-CMB  data are able to break parameter degeneracies from CMB measurements which results in being used as a robust geometrical test to impose constraints on the background evolution of modified gravities and dark energy  models \citep{Planck:2018vyg}. Since the sound speed before drag epoch depends only on the  ratio of the photon to baryon density, this sound horizon scale serves as a standard ruler and can be extracted from galaxy redshift surveys.

Information based on transverse measurements of galaxy redshift survey constrains the ratio of the comoving angular diameter distance  $D_{\rm M}$ and the sound horizon  $r_{\rm drag}$ at the drag epoch, $D_{M}/r_{\rm drag}$. Besides, information based on radial measurements yields $H(z)r_{\rm drag}$. BAO measurements from Baryon Oscillation Spectroscopic Survey (BOSS) Data Release 12 \citep{BOSS:2016wmc} provide measurements of both the Hubble parameter $H(z )r_{\rm drag}$ and the comoving angular diameter distance $D_{M} (z )/r_{\rm drag}$, at three correlated redshift bins $z_{\rm eff}$= 0.38, 0.51 and 0.61.
It is customary to combine these two observables and form direction-averaged quantity $D_{V}(z)/r_{\rm drag} = [cz D^2_{M}(z)H(z)^{-1}]^{1/3}/r_{\rm drag}$ based on the combination of transverse and radial BAO modes.  BAO measurements of 6-degree-Field Galaxy survey (6dFGS) \citep{2011MNRAS.416.3017B} and Sloan Digital Sky Survey (SDSS) Data Release 7 Main Galaxy Sample (SDSS-MGS) \citep{Ross:2014qpa} constrain the spherically averaged quantity $D_{V} (z)/r_{\rm drag}$  at the effective redshifts $z_{\rm eff} = 0.106$ and $0.15$, respectively.

All the above BAO measurements are limited to effective redshift less than unity. However, using quasars provides conditions to extend BAO measurements to redshifts greater than unity.  \cite{2018MNRAS.473.4773A} have  measured direction-averaged quantity $D_{V}$ at an effective redshift of $z_{\rm eff}=1.52$, and at even higher redshifts, BAOs have been measured in Lyman $\alpha$ spectra of quasars at the effective redshift $z_{\rm eff} = 2.35$ \citep{2019A&A...629A..86B}, both using a sample of quasars from the extended Baryon
Oscillation Survey (eBOSS).

Following \cite{Planck:2018vyg}, we  only use low-redshift BAOs, namely 6dFGS and SDSS-MGS measurements of $D_V /r_{\rm drag}$ alongside the final BOSS Data Release 12 anisotropic BAO measurements, and neglect high-redshift BAO measurements. Also, we do not apply BAO measurements of  BOSS CMASS \citep{BOSS:2013rlg}  as well as WiggleZ \citep{Kazin:2014qga} because their correlation with each other and with the final DR12  BAO measurements has not been very well quantified due to the partial overlapping of their volume.

\subsubsection*{Redshift space distortion}
Anisotropic clustering of galaxies in redshift space is induced by peculiar velocities. This effect is known as redshift space distortion (RSD) and can provide constraints on the growth rate of structure and the amplitude of the matter power spectrum \citep{Percival:2008sh}. RSDs are related to the time-time component of the metric perturbation  $\Psi$ through the relativistic Euler equation
\begin{equation}\label{Euler}
u^{\prime}_{i} + \frac{1}{\bar \rho + \bar p}\left[\partial_{i}\delta p - \partial_{j} \Pi^{j}_{i}\right] + \mathcal{H}u_{i} +\frac{{\bar p^{\prime}}}{\bar \rho + \bar p}u_{i} +\partial_{i}\Psi = 0,
\end{equation}
which is space part of energy-momentum conservation $\nabla_{\mu} T^{\mu}_{~i}=0$. Here, the second term represents the gradient of stress energy while redshifting is encoded by $\mathcal{H}u_i$, in which $u_i$ is the velocity field. This equation  breaks degeneracy with gravitational lensing and ISW effect due to their sensitivity to the combination $\Phi + \Psi$.
Measurements of RSDs are usually described as constraints on the $f(z)\sigma_{8}(z)$.  Here, $f(a)={\rm d}\ln D(a)/{\rm d}\ln a$ scales the amplitude of linear matter growth $D(a)\equiv\delta_{m}(a)/\delta_{m}(1)$ with respect to the cosmic time $t$, which gives rise to  sensitive  constraints on the dark energy.

Since  RSDs are related to the scales where non-linear effects and galaxy bias are significant and must be correctly modeled, measuring $f(z)\sigma_{8}(z)$ is considerably more complicated than estimating the BAO scale from galaxy redshift surveys \citep{BOSS:2016wmc,Planck:2018vyg}. Nevertheless, new high-precision measurements from BOSS Data Release 12, provide the strongest constraint on RSDs with respect to other measurements.  In this work, BOSS DR12 measurements of the quantity $f(z)\sigma_{8}(z)$ at the aforementioned three redshift bins are used to employ full covariance, between these three RSD measurements and those of BAO quantities $H(z)r_{\rm drag}$
and $D_{\rm V} (z)/r_{\rm drag}$ at correlated redshift bins $z_{\rm eff}= \{0.38,0.51,0.61\}$.
\vspace*{-.3cm}
\subsubsection*{Type Ia supernovae}
Type Ia supernovae (SN) as standard candles  which provide luminosity distances up to redshift of unity and beyond are useful to impose constraints  on the expansion history of the Universe. The SN data have little statistical power compared to $Planck$ and BAO. Therefore, their main usage is to test dynamical dark energy models and modified gravities \citep{Planck:2018vyg}. The absolute luminosity of SN remains uncertain and is marginalized out, which leads to  an unconstrained $H_0$. But, since SN makes higher precision measurements of relative distance at lower redshift,  SN data are useful to explore the background cosmology at low redshifts, because BAO do not represent high precision constraints. This is due to the fact that an absolute scale which is provided by BAO relays on higher redshift and especially to the CMB acoustic scale at the drag epoch $z_{\rm drag} \cong 1060$ \citep{BOSS:2016wmc}.

Nevertheless,  the combination of SN objects with BAO measurements is remarkably powerful to  impose constraints on the low-redshift distance scale \citep{mehta,2014Anderson}.
We use the analysis of the Pantheon sample \citep{2018ApJ...859..101S}, including 279 SN Ia from the Pan-STARRS Medium Deep Survey in redshift $0.03 < z < 0.68$ alongside SN Ia from  SDSS, Supernova Legacy Survey, and Hubble Space Telescope  samples. The final Pantheon catalog includes 1048 objects, out to $z = 2.26$.

Our vector parameter is a 12-$D$ space which includes 10 cosmic  parameters $\mathcal{P}=\{\Omega_m,\Omega_b,\Omega_{\nu}h^2,n_s,\tau,A_s,h,w_0,w_a,w_b\}$ and 2 nuisance parameters $\mathcal{N}=\{A_{Planck},m\}$, where $m$ denotes the SN Ia absolute magnitude. Following \cite{2022MNRAS.tmp.2714L}, we perform ellipsoidal nested sampling $\textsc{MultiNest}$ with conditions $\tt live_{-}points$=25$*D$=300, $\tt {efficiency}$= 0.008, $\tt tolerance$= 0.1, and $\tt constant_{-}efficency$= $F$ for all considered models, using $Planck$ CMB anisotropies measurements, non-CMB data set, and  combination of $Planck$ and non-CMB probes.
We use $\textsc{ChainConsumer}$\footnote{\url{https://github.com/Samreay/ChainConsumer}}
 \citep{Hinton2016}
to plot and analyze chains.

\section{results}\label{result}
We impose observational constraints on  some well-known time-dependent dark energy models.  We generalize each model to the equation of state with three free parameters to find a limitation on the number of essential parameters.

\subsection{CPL Parametrization}
\begin{figure}
\includegraphics[width=\linewidth]{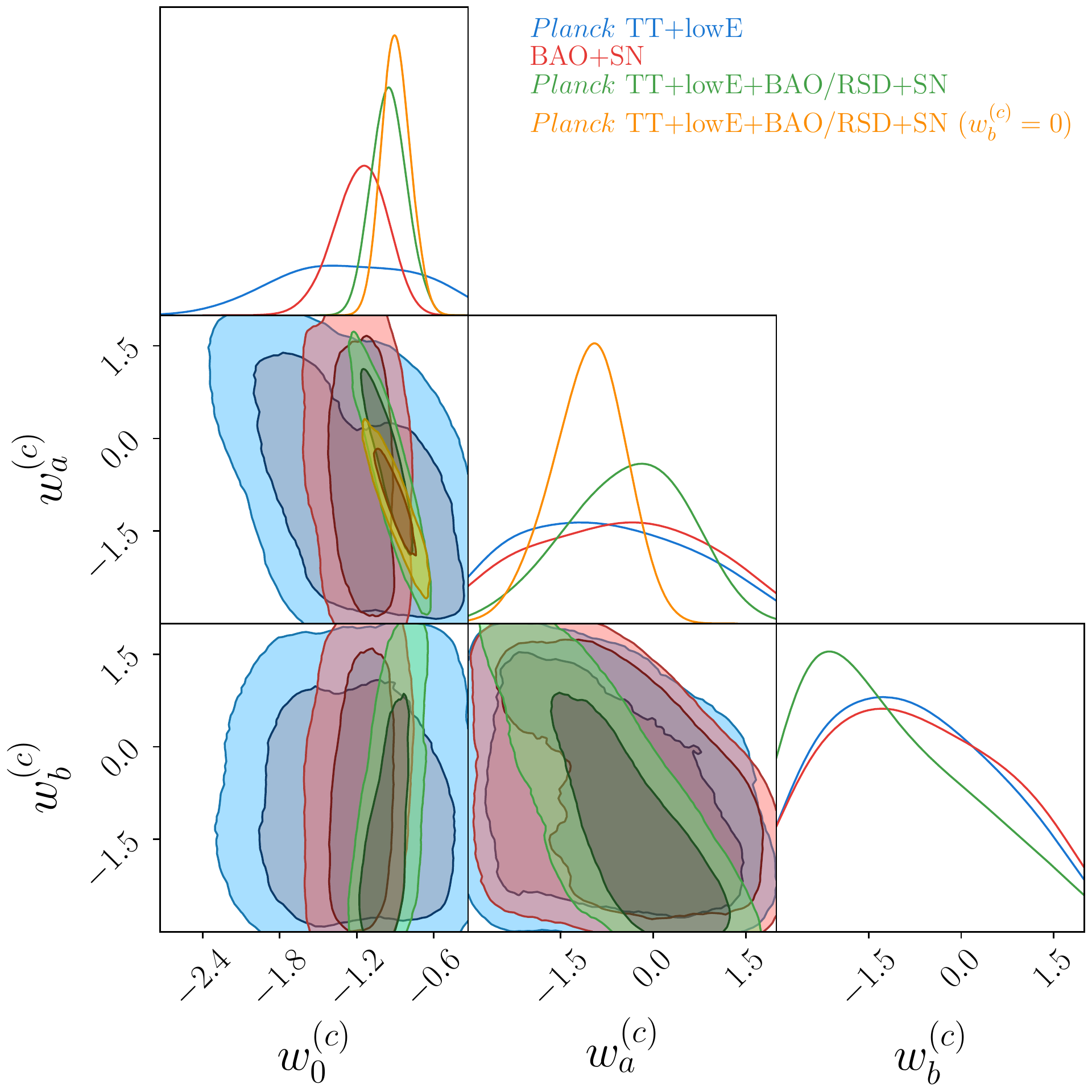}
\caption{Marginalized posterior distributions for the ($w^{(c)}_0 , w^{(c)}_a,w^{(c)}_b$) parameters of  extended CPL model $w(a)_{\rm cpl} = w^{(c)}_{0} + w^{(c)}_{a}(1-a)+w^{(c)}_b(1-a)^2$. This model does not provide a good fit to the $Planck$ TT+lowE, individually, or in combination with external data. However, the tightest constraints come from the combination $Planck$ TT+lowE+BAO/RSD+SN. The best and tightest constraint on $w^{(c)}_0$ and $w^{(c)}_a$ is obtained when $w^{(c)}_b=0$, indicated with orange contour.}
\label{fig1}
\end{figure}
The most trivial generalization for dark energy with a constant equation of state $w_0$ is
\begin{equation}\label{cpl1}
w(a)_{\rm cpl} = w^{(c)}_{0} + w^{(c)}_{a}(1-a),
\end{equation}
where in the literature is known as Chevallier--Polarski--Linder (CPL) parameterization \citep{Chevallier:2000qy,Linder:2002et}. At the high redshift equation of state is $w^{(c)}_0+w^{(c)}_a$ and at the low redshift becomes $w^{(c)}_0$, so $w^{(c)}_0$ is anticorrelated with $w^{(c)}_a$.
The simple generalization of CPL equation of state is
\begin{equation}\label{cpl2}
w(a)_{\rm cpl} = w^{(c)}_{0} + w^{(c)}_{a}(1-a) + w^{(c)}_{b} (1-a)^2.
\end{equation}
We use priors $-3 \le w^{(c)}_0 \le -0.33$, $-3 \le w^{(c)}_a \le 2$, and $-3 \le w^{(c)}_b \le 2$ alongside hard prior $w^{(c)}_0+w^{(c)}_a+w^{(c)}_b \le 0$ to demand that equation of state remains negative. Marginalized contour of the posterior distributions for $(w^{(c)}_0,w^{(c)}_a,w^{(c)}_b)$ are shown in Fig. \ref{fig1}. A wide volume of parameter space for the equation of state is allowed for both $Planck$ and external data. Adding external data to $Planck$ provides almost the tightest constraints.
We obtain
\begin{equation}
\left.\begin{aligned}
w^{(c)}_0 &= -0.95\pm 0.14, \\
w^{(c)}_a &= -0.18^{+0.88}_{-1.17}, \\
w^{(c)}_b &= -2.14^{+1.78}_{-0.78},
\end{aligned}\ \right\}\ \
\mbox{\text{\parbox{4.2cm}{\begin{flushleft}(68\,\% CL, $Planck$ TT+lowE\\+BAO/RSD+SN),\end{flushleft}}}}
\label{w0-wa-wb-cpl}
\end{equation}
which means the parameters $(w^{(c)}_0,w^{(c)}_a,w^{(c)}_b)$ lie  $0.3$, $0.2$ and $1.2\sigma$  away from $(-1,0,0)$.
Constraints on both $w^{(c)}_a$ and $w^{(c)}_b$ are poor from each data set. $Planck$ TT+lowE make $w^{(c)}_a$ and $w^{(c)}_b$ lie $0.7\sigma$ and $0.8\sigma$ away from zero. When external data are added,  the mean values of $w^{(c)}_a$ and $w^{(c)}_b$ change remarkably, and error bars of $w^{(c)}_a$ and $w^{(c)}_b$ are respectively reduced by $48\%$ and $45\%$. Fixing  $w^{(c)}_b=0$, we obtain the  constraints
\begin{equation}
\left.\begin{aligned}
w^{(c)}_0 &=-0.91^{+0.12}_{-0.11}, \\
w^{(c)}_a &=-0.94^{+0.55}_{-0.66},
\end{aligned}\ \right\}\ \
\mbox{\text{\parbox{4.2cm}{\begin{flushleft}(68\,\% CL, $Planck$ TT+lowE\\+BAO/RSD+SN),\end{flushleft}}}}
\label{w0-wa-cpl}
\end{equation}
which imply that $w^{(c)}_0$ and $w^{(c)}_a$ lie  $0.9$ and $1.7\sigma$ away from $-1$ and zero, respectively. Error bars of $w^{(c)}_a$ are decreased by $55\%$ and $78\%$ for the upper and lower bound, respectively.
\cite{SDSS:2004kqt} obtained $w^{(c)}_0=-0.98\pm0.19$ and $w^{(c)}_a={-0.05}^{+0.83}_{-0.65}$  of CPL parameterization for 1st year WMAP+SN measurements in combination with Ly$\alpha$ forest analysis of SDSS. Their results show that CPL is not different from $\Lambda$CDM. We find that at the $1\sigma$ confidence level,  the CPL model  is not similar  to $\Lambda$CDM cosmology for combination $Planck$ TT+lowE+BAO/RSD+SN.

The Bayes factor $\ln B$ is $-0.8$  and  $-0.6$  for CPL and extended form $w(a)= w_0 + w_a (1-a) + w_b (1-a)^2$ with respect to $\Lambda$CDM  for full data, respectively. These values for the Bayes factor imply that there is no significant preference for $\Lambda$ dark energy  over CPL and its extended form. 

\subsection{PADE Parametrization}\label{pade-sec}
\begin{figure}
\includegraphics[width=\linewidth]{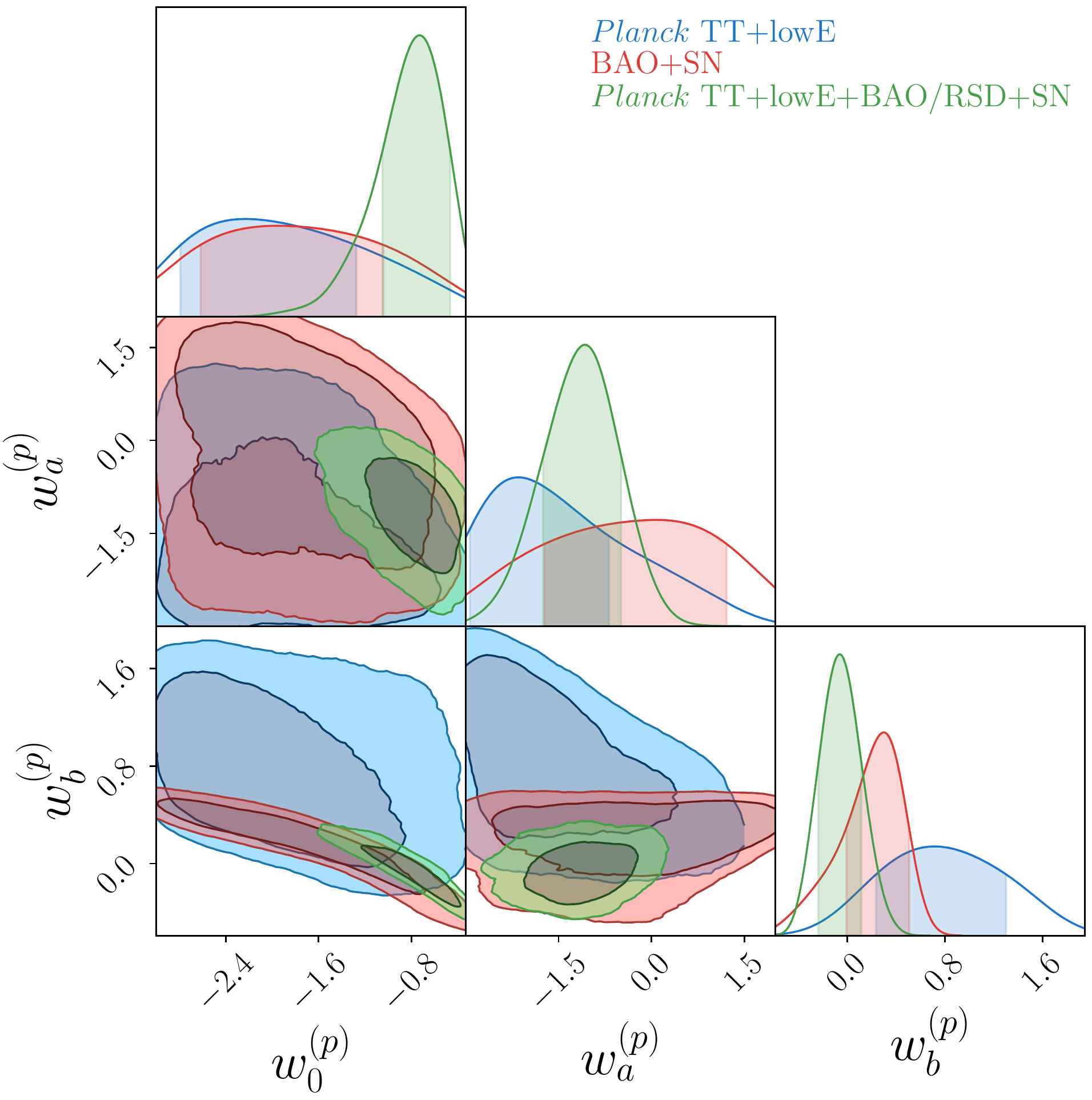}
\caption{Marginalized posterior distributions for ($w^{(p)}_0 , w^{(p)}_a,w^{(p)}_b$) parameters of the PADE model $w(a)_{\rm pade} = {[w^{(p)}_{0}+w^{(p)}_{a}(1-a)]}/{[1+w^{(p)}_{b}(1-a)]}$. $Planck$ TT+lowE provides poor constraints on the dark energy parameters.
The $R$ statistic with value $R=0.015$ implies that the $Planck$ CMB and geometrical data  are discordant when applied to PADE model. Therefore, the constraints given by equation (\ref{w0-wa-wb-pade}) are not reliable because two data sets must not be combined.}
\label{fig2}
\end{figure}

Instead of expressing equation of state $w(a)_{\rm cpl}=w^{(c)}_0+w^{(c)}_a(1-a)$ up to quadratic expansion, one can parameterize it as
\begin{equation}\label{pade}
w(a)_{\rm pade} = \frac{w^{(p)}_{0}+w^{(p)}_{a}(1-a)}{1+w^{(p)}_{b}(1-a)},
\end{equation}
where is named PADE parameterization \citep{Rezaei:2017yyj}. At the high redshifts equation of state is $(w^{(p)}_{0}+w^{(p)}_{ a})/(1+w^{(p)}_{ b})$, and at the low redshifts turns to $w^{(p)}_{0}$. Therefore, $w^{(p)}_{0}$ is anticorrelated with  both $w^{(p)}_{a}$ and $w^{(p)}_{b}$. Also, $w^{(p)}_{ a}$ and $w^{(p)}_{ b}$ are positively correlated. We use priors $-3\le w^{(p)}_0 \le -0.33$ and $-3 \le w^{(p)}_a \le 2$ with $-1 \le w^{(p)}_b \le 2$\break alongside  hard prior $w^{(p)}_0+w^{(p)}_a \le 0$ to ensure that equation of state remains non-singular and negative.
Marginalized contour of the posterior distributions for $(w^{(p)}_0,w^{(p)}_a,w^{(p)}_b)$ are shown in Fig. \ref{fig2}.
We find
\begin{equation}
\left.\begin{aligned}
w^{(p)}_0 &= -0.73^{+0.26}_{-0.32}, \\
w^{(p)}_a &= -1.07^{+0.58}_{-0.68}, \\
w^{(p)}_b &= -0.06\pm 0.18,
\end{aligned}\ \right\}\ \
\mbox{\text{\parbox{4.2cm}{\begin{flushleft}(68\,\% CL, $Planck$ TT+lowE\\+BAO/RSD+SN),\end{flushleft}}}}
\label{w0-wa-wb-pade}
\end{equation}
which shows the parameters $(w^{(p)}_0,w^{(p)}_a,w^{(p)}_b)$ lie $0.8$, $1.8$ and $0.3\sigma$  away from $(-1,0,0)$ and implies that PADE parameterization at the $1\sigma$ confidence level behaves like CPL parameterization in favour of full data.
The parameter $w^{(p)}_b$ imposes poor constraints on the $w^{(p)}_0$ with respect to $w^{(c)}_{0}$ and pushes $w^{(p)}_a$ somewhat to lower value in comparison with $w^{(c)}_{a}$ without  change on the error bars (see equation (\ref{w0-wa-cpl})).

Fig. \ref{fig2} implies that there could exist a tension between CMB and non-CMB data sets, which means one must care to combine them for PADE model. In Bayesian inference, the  $R$ statistic provides a measure for clarification tension between two different data sets \citep{2006PhRvD..73f7302M}. Given two different independent data sets $A$ and $B$, the Bayesian ratio $R$ is defined via
\begin{equation}
R = \frac{\mathcal{Z}_{A+B}}{\mathcal{Z}_{A}\mathcal{Z}_{B}},
\end{equation}
where evidence or marginal likelihood $\mathcal{Z}_{D}$, defined as the probability of measuring the observed data $D$ for a given model $\mathcal{M}$ is given by
\begin{equation}
\mathcal{Z}_{D}\equiv P(D|\mathcal{M})=\int {P(D|\theta,\mathcal{M})~P(\theta|\mathcal{M})~d\theta}= \int{\mathcal{L}_{D}~\pi~ d\theta}.
\end{equation}
Here, $\mathcal{L}_{D}\equiv P(D|\theta,\mathcal{M})$ is the likelihood  and $\pi\equiv P(\theta|\mathcal{M})$ denotes prior of parameters $\theta$. The value $R\gg1$  implies that both data sets are in agreement, while $R\ll1$ means that data sets are discordant. The $R$  statistic depends on the volume of prior $\pi$, and the tension can be hidden by increasing the prior width \citep{2021MNRAS.505.6179L}. Nevertheless,  if  $R$ indicates that two data sets are discordant, it should be taken seriously, because increasing the prior width  pushes $R$ to the larger value which means two data sets are in agreement \citep{2019PhRvD.100d3504H}.
For these roughly wide priors,  $R$ is
\begin{equation}
\ln R = \ln \mathcal{Z}_{Planck+ ext} - \ln \mathcal{Z}_{Planck}-\ln \mathcal{Z}_{ext} = - 4.17,
\end{equation}
where ``$ext$'' denotes geometrical data\footnote{$\ln \mathcal{Z}_{Planck+ ext}=-$1044.66,  $\ln \mathcal{Z}_{Planck}=-$505.05, and $\ln \mathcal{Z}_{ext}=-$535.44.}.
The value $R=0.015$ indicates that PADE dark energy model gives rise to a tension between $Planck$  and geometrical data. Thus, it is better that the PADE model not be considered a reliable parameterization. This tension indicates this model is not favoured by the data.

\begin{figure}
    \includegraphics[width=\linewidth]{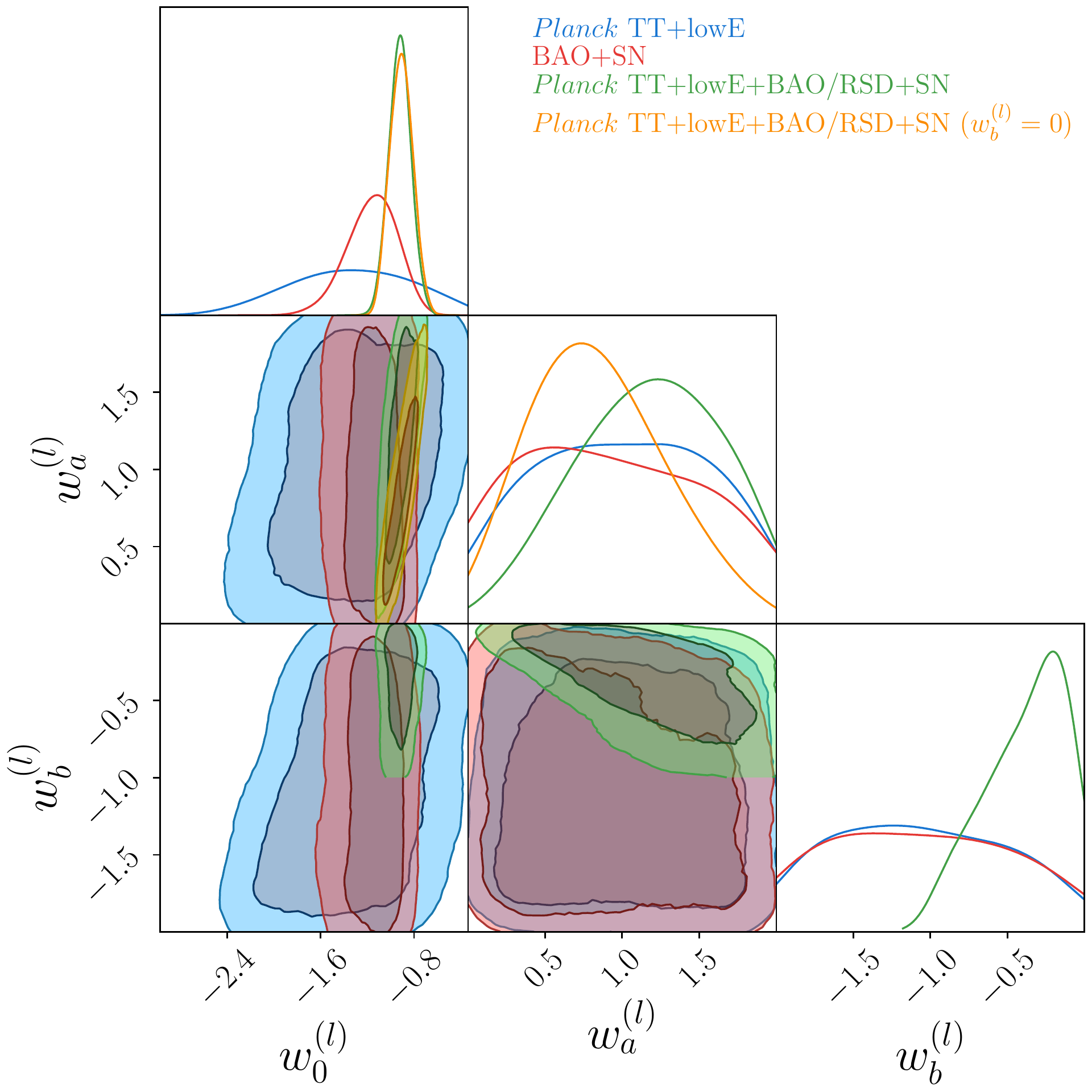}
    \caption{Marginalized posterior distributions of the ($w^{(l)}_0 , w^{(l)}_a,w^{(l)}_b$) parameters for logarithmic model $w(a)_{\rm log} = [{w^{(l)}_{0} + w^{(l)}_{a} \ln a}]/[{1+ w^{(l)}_{b} \ln a}]$ from various data. Using $Planck$ TT+lowE alone provides poor constraints with a wide area in parameter space. The tightest constraints come from the combination $Planck$ TT+lowE+BAO/RSD+SN. The orange contour corresponds to $w^{(l)}_b=0$, but as we can see the constraints in the plane ($w^{(l)}_0$,$w^{(l)}_a$) do not change significantly.}
    \label{fig4}
\end{figure}

\subsection{Logarithmic Parametrization}
Since the time scale of the expansion history of the Universe is given by $H^{-1}={\left(d \ln a/dt\right)}^{-1}$, following \cite{Efstathiou:1999tm}, is worth considering the time scale evolution of dark energy  in units of the e-folding scale $\ln a$
\begin{equation}\label{log1}
w(a)_{\rm log} = w^{(l)}_{ 0} + w^{(l)}_{ a} \ln a,
\end{equation}
with extended form
\begin{equation}\label{log2}
w(a)_{\rm log} = \frac{w^{(l)}_{0} + w^{(l)}_{a} \ln a}{1+ w^{(l)}_{ b} \ln a}.
\end{equation}
In order that the equation of state remains negative and non-singular,  prior of the parameters $w^{(l)}_{a}$ and $w^{(l)}_{b}$ must be chosen positive and negative, respectively. Therefore, $w^{(l)}_{0}$ with both $w^{(l)}_{a}$ and $w^{(l)}_{b}$ is positively correlated, and ($w^{(l)}_{a}$,$w^{(l)}_{b}$) are anti-correlated. We apply priors $-3\le w^{(l)}_0\le-0.33$, $0\le w^{(l)}_a\le2$, and $-2\le w^{(l)}_b\le 0$.
Marginalized contour of the posterior distributions for $(w^{(l)}_0,w^{(l)}_a,w^{(l)}_b)$ are shown in Fig. \ref{fig4}. A wide volume of parameter space for the equation of state is allowed for both $Planck$ and non-CMB data. Geometrical data provides tighter constraints than $Planck$ for $w^{(l)}_0$, but both data set impose poor constraints on $w^{(l)}_a$ and $w^{(l)}_b$. Combining these two data sets provides the tightest constraints with the values\footnote{To avoid  posterior probability distribution with two or more maxima, we restrict $w^{(l)}_b$ prior to $-1\le w^{(l)}_b\le 0$.}
\begin{equation}
\left.\begin{aligned}
w^{(l)}_0 &= -0.92^{+0.09}_{-0.10}, \\
w^{(l)}_a &= 1.23 \pm 0.50, \\
w^{(l)}_b &= -0.20^{+0.17}_{-0.35},
\end{aligned}\ \right\}\ \
\mbox{\text{\parbox{4.2cm}{\begin{flushleft}(68\,\% CL, $Planck$ TT+lowE\\+BAO/RSD+SN).\end{flushleft}}}}
\label{w0-wa-wb-log}
\end{equation}
Dark energy parameters $(w^{(l)}_0,w^{(l)}_a,w^{(l)}_b)$ lie $1$, $2.8$ and $1.1\sigma$ away from $\Lambda$CDM values.
Fixing the parameter $w^{(l)}_b=0$, we obtain the  constraints
\begin{equation}
\left.\begin{aligned}
w^{(l)}_0 &= -0.91 \pm 0.10, \\
w^{(l)}_a &= 0.74^{+0.48}_{-0.42},
\end{aligned}\ \right\}\ \
\mbox{\text{\parbox{4.2cm}{\begin{flushleft}(68\,\% CL, $Planck$ TT+lowE\\+BAO/RSD+SN),\end{flushleft}}}}
\label{w0-wa-log}
\end{equation}
which implies $w^{(l)}_0$ and $w^{(l)}_a$ lie  $0.9$ and $1.7\sigma$ away from $-1$ and zero, respectively.
The parameter $w^{(l)}_b$ does not change significantly constraints on $w^{(l)}_0$, but  pushes the peak of  $w^{(l)}_a$  to larger value without significant change on the error bars. \cite{2022A&A...668A.135S} obtained $w^{(l)}_0=-1.18 \pm 0.11$ and $w^{(l)}_a = -0.20\pm0.86$ which means $w^{(l)}_a$ lies $0.2\sigma$ away from zero for BAO+SN.

The Bayes factor $\ln B$ is $0.2$ and $-0.8$ for $(w^{(l)}_0,w^{(l)}_a)$ and $(w^{(l)}_0,w^{(l)}_a,w^{(l)}_b)$ models with respect to $\Lambda$CDM cosmology for $Planck$ TT+lowE+BAO/RSD+SN, respectively.
The values $\ln B=0.2$ and $-0.8$ indicate that the Bayes factor  is inconclusive in terms of
preference for $w(a)=w_0+w_a\ln a$ and $\Lambda$ dark energy, respectively.

\subsection{Oscillating Parameterization}
\begin{figure}
\includegraphics[width=\linewidth]{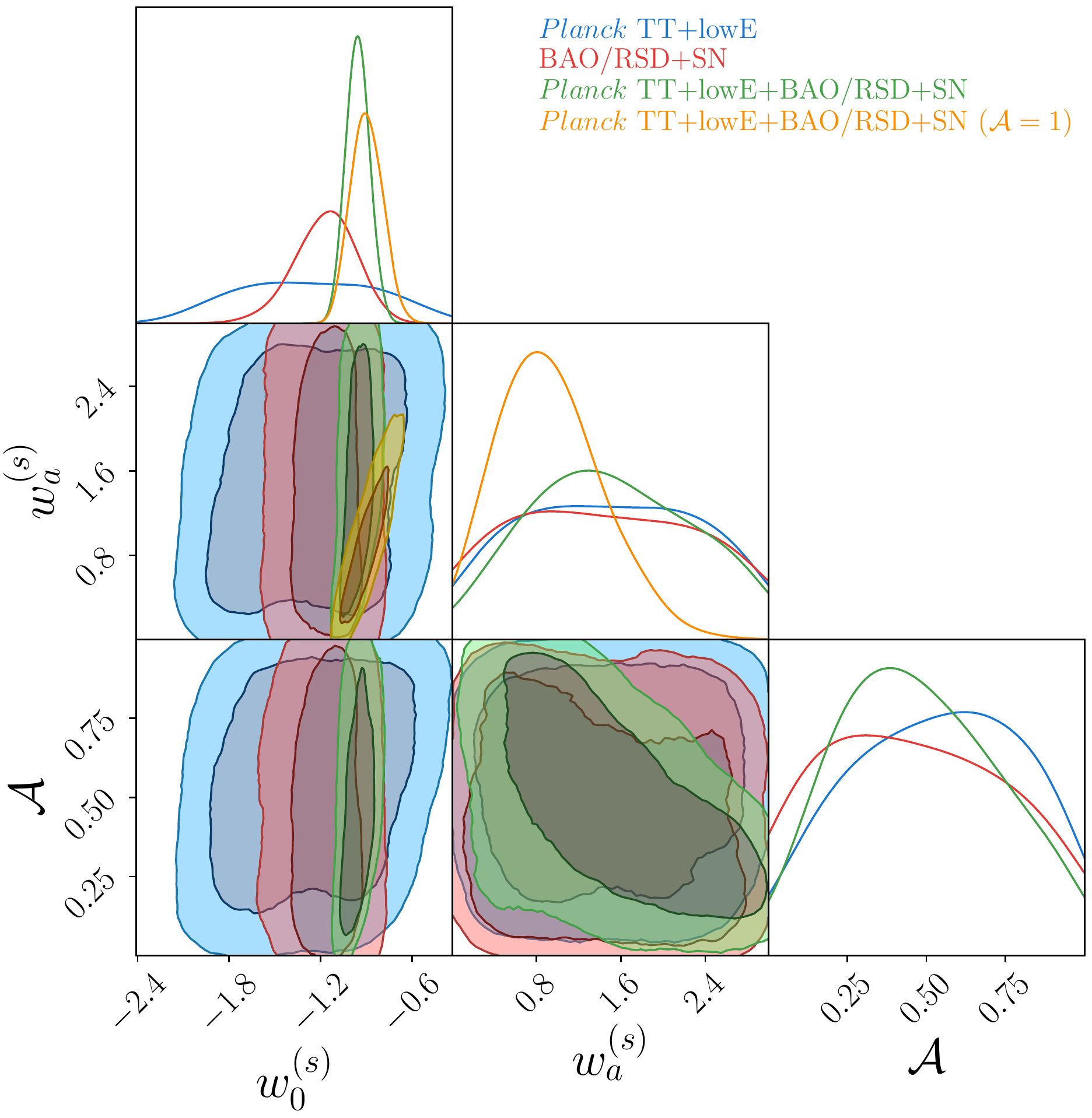}
\caption{Marginalized posterior distributions of the ($w^{(s)}_0 , w^{(s)}_a,\mathcal{A}$) parameters for oscillating parameterization $w(a)_{\rm sin} = w^{(s)}_{0} + w^{(s)}_{ a} \sin (\mathcal{A}\ln a)$ from various data. The tightest constraints come from the combination $Planck$ TT+lowE+BAO/RSD+SN. Using $Planck$ TT+lowE alone provides poor constraints with a wide area in parameter space that corresponds to large values of the Hubble parameter. The orange contour corresponds to the two free parameters model with $\mathcal{A}=1$, which obviously shows the best and tightest constraints on the $w^{(s)}_a$.}
\label{fig5}
\end{figure}

\begin{table*}
    \begin{center}
        \begin{tabular}{ccccccc}
            \noalign{\hrule\vskip 1pt}
            \noalign{\hrule\vskip 1pt}
            Number of & Dark energy & Parameters &$\ln B_{w(a)/{\rm{\Lambda}}}$& Outcome\\
            parameters & model &  &$Planck$ TT+lowE+BAO/RSD+SN&\\
            \noalign{\hrule\vskip 2pt}
            \multirow{2}{*}{}
            $1$& $\Lambda$ & $w_\Lambda=-1$ & $~~~\cdots$ & $\cdots$
            \vspace{.1mm}
            \\
            \noalign{\hrule\vskip 1pt}
            & $w_{\rm 0}+w_{ a}(1-a)$ & $[w_{\rm 0},w_{ a}]$  & $ -0.8$ & The preference for $\Lambda$ is inconclusive
            \vspace{1mm}
            \\
            2 & $w_{\rm 0}+w_{ a} \ln a$ & $[w_{\rm 0},w_{ a}]$ & $\textcolor{white}{+}0.2$  & The preference for $w(a)$ model is inconclusive
            \vspace{1mm}
            \\
            & $w_{\rm 0}+w_{ a} \sin (\ln a)$ & $ [w_{\rm 0},w_{ a}]$& $-2.0$ & The preference for  $\Lambda$ is moderate
            \vspace{1mm}
            \\
            \noalign{\hrule\vskip 2pt}
            & $w_{\rm 0}+w_{ a}(1-a)+w_{ b}(1-a)^2$ & $[w_{\rm 0},w_{ a},w_{ b}]$  & $-0.6$ & The preference for $\Lambda$ is inconclusive
            \vspace{1mm}
            \\
            & $\left[{w_{\rm 0}+w_{ a}(1-a)}\right]\big/{\left[1+w_{ b}(1-a)\right]}$ & $[w_{\rm 0},w_{ a},w_{ b}]$  &
            $-2.6~ (\rm  unreliable)$  & The $w(a)$ model cannot be compared due to tension
            \vspace{1mm}
            \\
            3& $({w_{\rm 0}+w_{ a} \ln a})\big/({1+w_{ b}\ln a})$ & $[w_{\rm 0},w_{ a},w_{ b}]$ & $-0.8$ &  The preference for $\Lambda$ is inconclusive
            \vspace{1mm}
            \\
            & $w_{\rm 0}+w_{ a} \sin ({\mathcal A}\ln a)$ & $ [w_{\rm 0},w_{ a},\mathcal A]$ & $-1.1$ & The $w(a)$ model is weakly unsupported
            \vspace{1mm}
            \\
            \noalign{\hrule\vskip 2pt}
        \end{tabular}
    \end{center}
    \caption{Bayesian comparison for time-dependent dark energy models from $Planck$  in combination with  external data.  Here, the Bayes factor is described as
$\ln B_{w(a)/{\rm{\Lambda}}}=\ln{(\mathcal{Z}_{w(a)}/\rm \mathcal{Z}_{{\Lambda}})}$ with $0.4$ as the error on the $\ln B_{w(a)/{\rm{\Lambda}}}$.
A negative (positive) value for $\ln B_{w(a)/{\rm{\Lambda}}}$ indicates that the competing model is unsupported (supported) with respect to the $\Lambda$CDM model.        
For both $w(a)=w_0 +w_a(1-a)$ and $w(a)=w_0 + w_a (1-a) + w_b (1-a)^2$, the Bayes factor is  inconclusive in terms of the preference for $\Lambda$. Nevertheless, the former is the most favourable model because the presence of $w_b$ leads to poor constraints on the parameters $w_0$ and $w_a$.
There is no significant support  in favour of  dark energy $w(a)=w_0+w_a\ln a$ over $\Lambda$ because this preference is inconclusive. For the equation of state $w(a)=(w_0+w_a\ln a)/(1+w_b \ln a)$, the Bayes factor is inconclusive in terms of preference for $\Lambda$. The model $w(a)=w_0 + w_a \sin (\ln a)$ has  the least strength of support where $\Lambda$ dark energy is moderately supported,  and the equation of state $w(a)=w_0 + w_a \sin (\mathcal{A}\ln a)$ is weakly unsupported. The Bayesian evidence estimation for  model $w(a)=\left[{w_{\rm 0}+w_{ a}(1-a)}\right]\big/{\left[1+w_{ b}(1-a)\right]}$ is not reliable as this model gives rise to a tension between $Planck$ and BAO/RSD+SN when applied for this model, hence, this model cannot be considered as a reliable model.}
\label{compare}
\end{table*}
An oscillating dark energy equation of state can potentially solve coincidence problem \citep{2000PhRvL..85.5276D,Nojiri:2006ww}. It is worth mentioning that there is no need to have oscillating potential $V(\varphi)$ for investigating oscillation behavior in the dark energy equation of state $w(a)$ \citep{Linder2006OnOD}.
Periodicity of dark energy is described in terms of $\ln a$
\begin{equation}\label{sin}
w(a)_{\rm sin} = w^{(s)}_{0} +w^{(s)}_{ a} \sin (\mathcal{A}\ln a + \theta),
\end{equation}
because $H^{-1}=(d\ln a/dt)^{-1}$ represents natural period of the cosmic expansion. For the sine function to be injective, we set  phase  $\theta = 0$ to have $-\pi/2 \le \mathcal A \ln a \le 0$
for a wide range of values of the scale factor $a$.
The prior $0\le \mathcal A \le1$ provides the condition that the sine function is injective for scale factor $a \ll 1$.
With a positive $\mathcal A$, $w^{(s)}_{0}$ is positively correlated with $w^{(s)}_{a}$ and $\mathcal A$. We apply prior $-3\le w^{(s)}_0\le-0.33$, $0\le w^{(s)}_a\le3$  and $0 \le \mathcal A \le 1$.
For $\mathcal A=1$, we obtain the constraints
\begin{equation}
\left.\begin{aligned}
w^{(s)}_0 &= -0.91^{+0.12}_{-0.10}, \\
w^{(s)}_a &= 0.80^{+0.53}_{-0.47},
\end{aligned}\ \right\}\ \
\mbox{\text{\parbox{4.2cm}{\begin{flushleft}(68\,\% CL, $Planck$ TT+lowE\\+BAO/RSD+SN),\end{flushleft}}}}
\label{w0-wa-osc}
\end{equation}
which indicates that $w^{(s)}_0$ and $w^{(s)}_a$ lie $0.9$ and $1.7\sigma$ away  from $\Lambda$CDM values, respectively.
\cite{2018PhRvD..98f3510P} obtained $w^{(s)}_0 = -0.98 \pm 0.05$ and $w^{(s)}_a = 0.01\pm 0.03$
for $Planck$ (2015)+BAO/RSD+SN. These results show that  this model is similar to $\Lambda$CDM because $w^{(s)}_0$ and $w^{(s)}_a$ are different from -1 and zero at the $0.4$ and $0.3\sigma$ confidence level, respectively. For $\mathcal A\neq1$, $Planck$ data shift  the peak of the posterior distribution of $\mathcal A$ toward zero value. Adding external data changes significantly  constraints on the $w^{(s)}_0$ but has no major impact on the posterior of both $w^{(s)}_a$ and $\mathcal A$.  For the full data set, we obtain
\begin{equation}
\left.\begin{aligned}
w^{(s)}_0 &= -0.95^{+0.07}_{-0.08}, \\
w^{(s)}_a &= 1.29^{+0.97}_{-0.69}, \\
\mathcal A &= 0.38^{+0.32}_{-0.21},
\end{aligned}\ \right\}\ \
\mbox{\text{\parbox{4.2cm}{\begin{flushleft}(68\,\% CL, $Planck$ TT+lowE\\+BAO/RSD+SN),\end{flushleft}}}}
\label{w0-wa-wb-osc}
\end{equation}
which means  dark energy parameters $(w^{(s)}_0,w^{(s)}_a,\mathcal A)$ lie $0.6$, $1.9$ and $1.9\sigma$ away from $(-1,0,1)$. Marginalized contour of the posterior distributions for $(w^{(o)}_0,w^{(o)}_a,\mathcal A)$ are shown in Fig. \ref{fig5}. A varying amplitude $\mathcal A$ changes significantly constraints on $w^{(s)}_0$ and $w^{(s)}_a$. The constraints on $w^{(s)}_0$ get somewhat improved and the center is pushed to -1. The peak of posterior distributions of $w^{(o)}_a$ moves toward to larger value and error bars are increased by $80\%$ for the upper band for $Planck$ TT+lowE+BAO/RSD+SN. When we only apply $Planck$ measurements, the Hubble parameter $H_0$ turns to the high value corresponding to the phantom region as a result of priors of the equation of state parameters.

The Bayes factor $\ln B$ is $-2.0$ and $-1.1$ for
$(w^{(s)}_0,w^{(s)}_a)$ and $(w^{(s)}_0,w^{(s)}_a,\mathcal A)$
models with respect to $\Lambda$CDM cosmology for $Planck$
TT+lowE+BAO/RSD+SN, respectively. These results show that
the strength of support for  $\Lambda$ dark energy over
$w(a)=w_0+w_a \sin(\ln a)$ is moderate,  and the 
preference for $\Lambda$ over equation of state $w(a)=w_0+w_a \sin(\mathcal{A}\ln
a)$ is weak.

\section{Discussion \& Conclusion}\label{discuss}
In this work, we imposed observational constraints and performed a model selection approach on the different models of time-dependent dark energy models  in favour of  $Planck$ temperature and polarization data alongside probes based on distance measurements in combination with RSD measurements.
We considered models with two free parameters and then generalize each model to three free parameters in the equation of state.
For all models, $Planck$ TT+lowE and non-CMB probes separately provide poor constraints on the parameters and combining them gives rise to the tightest posterior distributions.

Instead of expanding scale factor up to quadratic term $(1-a)^2$, we used PADE model $w(a)=[w_0+w_a(1-a)]/[1+w_b(1-a)]$. The $R$ statistic as tension metric indicates that the PADE model gives rise to a tension between $Planck$ CMB measurements and a combination of  BAO/RSD+SN which translates into a requirement that  there is a problem with the PADE parameterization and it cannot be considered as a reliable model.

We have summarized the Bayes factor for all considered models in Table \ref{compare} for full data, in which we quote $0.4$ as the error on the $\ln B_{w(a)/{\rm{\Lambda}}}$.
According to the values of the Bayes factor and considering the dispersion of posterior distributions of dark energy free parameters, two-parameter models are sufficient for investigating the time dependency behavior of  dark energy, compared to their three free parameter parameterizations. Among models with two free parameters, only logarithmic model $w(a)=w_0+w_a \ln a$ has a preference over $\Lambda$ dark energy but this strength of support is not significant because $\ln B_{w(a)/{\rm{\Lambda}}}=0.2$.
The preference for $\Lambda$ over CPL model $w(a)=w_0+w_a(1-a)$ is inconclusive with  $\ln B_{w(a)/{\rm{\Lambda}}}=-0.8$, and oscillating dark energy $w(a)=w_0+w_a\sin(\ln a)$ is almost moderately unsupported with respect to $\rm \Lambda CDM$ with  $\ln B_{w(a)/{\rm{\Lambda}}}=-2.0$.

For the competing models with three free parameters, the Bayesian evidence implies that the strength of support for $\Lambda$ dark  energy over both equations of states  $w(a)=w_0+w_a(1-a)+w_b(1-a)^2$ and $w(a)=[w_0+w_a\ln a]/[1+w_b\ln a]$ is inconclusive with receptively  $\ln B_{w(a)/{\rm{\Lambda}}}=-0.6$ and $-0.8$, and time-dependent dark energy  $w(a)=w_0+w_a \sin(\mathcal A \ln a)$ is weakly unsupported with respect to $\rm \Lambda CDM$ with $\ln B_{w(a)/{\rm{\Lambda}}}=-1.1$.

\section{acknowledgment}
We deeply thank the referee for valuable comments and feedback.
The analysis for this paper was run on the Gavazang cluster of the
Institute for Advanced Studies in Basic Sciences (IASBS). Also, We are grateful to Bj$\ddot {\rm o}$rn Malte Sch$\ddot
{\rm a}$fer for constructive comments.  

\black
\section{Data Availability}
All data used in this work are publicly available and implemented
to public package  $\tt CosmoSIS$. $Planck$ CMB angular
anisotropies power spectra are available from Planck Legacy
Archive \url{https://pla.esac.esa.int/#cosmology}. The BAO
measurements are available from the cited papers, and the Pantheon
supernovae data sets are available from
\url{https://github.com/dscolnic/Pantheon}.

\bibliographystyle{mnras}
\bibliography{ref}

\begin{thebibliography}{}
\makeatletter
\relax
\def\mn@urlcharsother{\let\do\@makeother \do\$\do\&\do\#\do\^\do\_\do\%\do\~}
\def\mn@doi{\begingroup\mn@urlcharsother \@ifnextchar [ {\mn@doi@}
  {\mn@doi@[]}}
\def\mn@doi@[#1]#2{\def\@tempa{#1}\ifx\@tempa\@empty \href
  {http://dx.doi.org/#2} {doi:#2}\else \href {http://dx.doi.org/#2} {#1}\fi
  \endgroup}
\def\mn@eprint#1#2{\mn@eprint@#1:#2::\@nil}
\def\mn@eprint@arXiv#1{\href {http://arxiv.org/abs/#1} {{\tt arXiv:#1}}}
\def\mn@eprint@dblp#1{\href {http://dblp.uni-trier.de/rec/bibtex/#1.xml}
  {dblp:#1}}
\def\mn@eprint@#1:#2:#3:#4\@nil{\def\@tempa {#1}\def\@tempb {#2}\def\@tempc
  {#3}\ifx \@tempc \@empty \let \@tempc \@tempb \let \@tempb \@tempa \fi \ifx
  \@tempb \@empty \def\@tempb {arXiv}\fi \@ifundefined
  {mn@eprint@\@tempb}{\@tempb:\@tempc}{\expandafter \expandafter \csname
  mn@eprint@\@tempb\endcsname \expandafter{\@tempc}}}

\bibitem[\protect\citeauthoryear{{Alam} et~al.,}{{Alam}
  et~al.}{2017}]{BOSS:2016wmc}
{Alam} S.,  et~al., 2017, \mn@doi [\mnras] {10.1093/mnras/stx721}, \href
  {https://ui.adsabs.harvard.edu/abs/2017MNRAS.470.2617A} {470, 2617}

\bibitem[\protect\citeauthoryear{{Anderson} et~al.,}{{Anderson}
  et~al.}{2014a}]{2014Anderson}
{Anderson} L.,  et~al., 2014a, \mn@doi [\mnras] {10.1093/mnras/stt2206}, \href
  {https://ui.adsabs.harvard.edu/abs/2014MNRAS.439...83A} {439, 83}

\bibitem[\protect\citeauthoryear{{Anderson} et~al.,}{{Anderson}
  et~al.}{2014b}]{BOSS:2013rlg}
{Anderson} L.,  et~al., 2014b, \mn@doi [\mnras] {10.1093/mnras/stu523}, \href
  {https://ui.adsabs.harvard.edu/abs/2014MNRAS.441...24A} {441, 24}

\bibitem[\protect\citeauthoryear{{Ata} et~al.,}{{Ata}
  et~al.}{2018}]{2018MNRAS.473.4773A}
{Ata} M.,  et~al., 2018, \mn@doi [\mnras] {10.1093/mnras/stx2630}, \href
  {https://ui.adsabs.harvard.edu/abs/2018MNRAS.473.4773A} {473, 4773}

\bibitem[\protect\citeauthoryear{{Audren}, {Lesgourgues}, {Benabed}  \&
  {Prunet}}{{Audren} et~al.}{2013}]{Audren:2012wb}
{Audren} B.,  {Lesgourgues} J.,  {Benabed} K.,   {Prunet} S.,  2013, \mn@doi
  [\jcap] {10.1088/1475-7516/2013/02/001}, \href
  {https://ui.adsabs.harvard.edu/abs/2013JCAP...02..001A} {2013, 001}

\bibitem[\protect\citeauthoryear{{Ballesteros}, {Hollenstein}, {Jain}  \&
  {Kunz}}{{Ballesteros} et~al.}{2012}]{Ballesteros:2011cm}
{Ballesteros} G.,  {Hollenstein} L.,  {Jain} R.~K.,   {Kunz} M.,  2012, \mn@doi
  [\jcap] {10.1088/1475-7516/2012/05/038}, \href
  {https://ui.adsabs.harvard.edu/abs/2012JCAP...05..038B} {2012, 038}

\bibitem[\protect\citeauthoryear{{Bertschinger}}{{Bertschinger}}{2006}]{bertschinger2006growth}
{Bertschinger} E.,  2006, \mn@doi [\apj] {10.1086/506021}, \href
  {https://ui.adsabs.harvard.edu/abs/2006ApJ...648..797B} {648, 797}

\bibitem[\protect\citeauthoryear{{Beutler} et~al.,}{{Beutler}
  et~al.}{2011}]{2011MNRAS.416.3017B}
{Beutler} F.,  et~al., 2011, \mn@doi [\mnras]
  {10.1111/j.1365-2966.2011.19250.x}, \href
  {https://ui.adsabs.harvard.edu/abs/2011MNRAS.416.3017B} {416, 3017}

\bibitem[\protect\citeauthoryear{{Blas}, {Lesgourgues}  \& {Tram}}{{Blas}
  et~al.}{2011}]{Blas:2011rf}
{Blas} D.,  {Lesgourgues} J.,   {Tram} T.,  2011, \mn@doi [\jcap]
  {10.1088/1475-7516/2011/07/034}, \href
  {https://ui.adsabs.harvard.edu/abs/2011JCAP...07..034B} {2011, 034}

\bibitem[\protect\citeauthoryear{{Blomqvist} et~al.,}{{Blomqvist}
  et~al.}{2019}]{2019A&A...629A..86B}
{Blomqvist} M.,  et~al., 2019, \mn@doi [\aap] {10.1051/0004-6361/201935641},
  \href {https://ui.adsabs.harvard.edu/abs/2019A&A...629A..86B} {629, A86}

\bibitem[\protect\citeauthoryear{{Caldwell}}{{Caldwell}}{2002}]{caldwell2002phantom}
{Caldwell} R.~R.,  2002, \mn@doi [Physics Letters B]
  {10.1016/S0370-2693(02)02589-3}, \href
  {https://ui.adsabs.harvard.edu/abs/2002PhLB..545...23C} {545, 23}

\bibitem[\protect\citeauthoryear{{Caldwell}, {Dave}  \&
  {Steinhardt}}{{Caldwell} et~al.}{1998}]{Caldwell:1997ii}
{Caldwell} R.~R.,  {Dave} R.,   {Steinhardt} P.~J.,  1998, \mn@doi [\prl]
  {10.1103/PhysRevLett.80.1582}, \href
  {https://ui.adsabs.harvard.edu/abs/1998PhRvL..80.1582C} {80, 1582}

\bibitem[\protect\citeauthoryear{{Chevallier} \& {Polarski}}{{Chevallier} \&
  {Polarski}}{2001}]{Chevallier:2000qy}
{Chevallier} M.,  {Polarski} D.,  2001, \mn@doi [International Journal of
  Modern Physics D] {10.1142/S0218271801000822}, \href
  {https://ui.adsabs.harvard.edu/abs/2001IJMPD..10..213C} {10, 213}

\bibitem[\protect\citeauthoryear{{DES Collaboration}}{{DES
  Collaboration}}{2018}]{2018PhRvD..98d3526A}
{DES Collaboration} 2018, \mn@doi [\prd] {10.1103/PhysRevD.98.043526}, \href
  {https://ui.adsabs.harvard.edu/abs/2018PhRvD..98d3526A} {98, 043526}

\bibitem[\protect\citeauthoryear{{DES Collaboration}}{{DES
  Collaboration}}{2021}]{2021MNRAS.505.6179L}
{DES Collaboration} 2021, \mn@doi [\mnras] {10.1093/mnras/stab1670}, \href
  {https://ui.adsabs.harvard.edu/abs/2021MNRAS.505.6179L} {505, 6179}

\bibitem[\protect\citeauthoryear{{DES Collaboration}}{{DES
  Collaboration}}{2022}]{2022MNRAS.tmp.2714L}
{DES Collaboration} 2022, arXiv e-prints, \href
  {https://ui.adsabs.harvard.edu/abs/2022arXiv220208233L} {p. arXiv:2202.08233}

\bibitem[\protect\citeauthoryear{{Dodelson}, {Kaplinghat}  \&
  {Stewart}}{{Dodelson} et~al.}{2000}]{2000PhRvL..85.5276D}
{Dodelson} S.,  {Kaplinghat} M.,   {Stewart} E.,  2000, \mn@doi [\prl]
  {10.1103/PhysRevLett.85.5276}, \href
  {https://ui.adsabs.harvard.edu/abs/2000PhRvL..85.5276D} {85, 5276}

\bibitem[\protect\citeauthoryear{{Efstathiou}}{{Efstathiou}}{1999}]{Efstathiou:1999tm}
{Efstathiou} G.,  1999, \mn@doi [\mnras] {10.1046/j.1365-8711.1999.02997.x},
  \href {https://ui.adsabs.harvard.edu/abs/1999MNRAS.310..842E} {310, 842}

\bibitem[\protect\citeauthoryear{{Einstein}}{{Einstein}}{1917}]{1917SPAW.......142E}
{Einstein} A.,  1917, Sitzungsberichte der K{\"o}niglich Preu{\ss}ischen
  Akademie der Wissenschaften (Berlin), \href
  {https://ui.adsabs.harvard.edu/abs/1917SPAW.......142E} {pp 142--152}

\bibitem[\protect\citeauthoryear{{Eisenstein}, {Hu}  \& {Tegmark}}{{Eisenstein}
  et~al.}{1998}]{Eisenstein:1998tu}
{Eisenstein} D.~J.,  {Hu} W.,   {Tegmark} M.,  1998, \mn@doi [\apjl]
  {10.1086/311582}, \href
  {https://ui.adsabs.harvard.edu/abs/1998ApJ...504L..57E} {504, L57}

\bibitem[\protect\citeauthoryear{{Fang}, {Hu}  \& {Lewis}}{{Fang}
  et~al.}{2008}]{Fang:2008sn}
{Fang} W.,  {Hu} W.,   {Lewis} A.,  2008, \mn@doi [\prd]
  {10.1103/PhysRevD.78.087303}, \href
  {https://ui.adsabs.harvard.edu/abs/2008PhRvD..78h7303F} {78, 087303}

\bibitem[\protect\citeauthoryear{{Feroz}, {Hobson}  \& {Bridges}}{{Feroz}
  et~al.}{2009}]{Feroz:2008xx}
{Feroz} F.,  {Hobson} M.~P.,   {Bridges} M.,  2009, \mn@doi [\mnras]
  {10.1111/j.1365-2966.2009.14548.x}, \href
  {https://ui.adsabs.harvard.edu/abs/2009MNRAS.398.1601F} {398, 1601}

\bibitem[\protect\citeauthoryear{{Feroz}, {Hobson}, {Cameron}  \&
  {Pettitt}}{{Feroz} et~al.}{2019}]{Feroz:2013hea}
{Feroz} F.,  {Hobson} M.~P.,  {Cameron} E.,   {Pettitt} A.~N.,  2019, \mn@doi
  [The Open Journal of Astrophysics] {10.21105/astro.1306.2144}, \href
  {https://ui.adsabs.harvard.edu/abs/2019OJAp....2E..10F} {2, 10}

\bibitem[\protect\citeauthoryear{{Handley} \& {Lemos}}{{Handley} \&
  {Lemos}}{2019}]{2019PhRvD.100d3504H}
{Handley} W.,  {Lemos} P.,  2019, \mn@doi [\prd] {10.1103/PhysRevD.100.043504},
  \href {https://ui.adsabs.harvard.edu/abs/2019PhRvD.100d3504H} {100, 043504}

\bibitem[\protect\citeauthoryear{{Hinton}}{{Hinton}}{2016}]{Hinton2016}
{Hinton} S.~R.,  2016, \mn@doi [The Journal of Open Source Software]
  {10.21105/joss.00045}, \href
  {https://ui.adsabs.harvard.edu/abs/2016JOSS....1...45H} {1, 00045}

\bibitem[\protect\citeauthoryear{{Hu}}{{Hu}}{2005}]{Hu:2004kh}
{Hu} W.,  2005, \mn@doi [\prd] {10.1103/PhysRevD.71.047301}, \href
  {https://ui.adsabs.harvard.edu/abs/2005PhRvD..71d7301H} {71, 047301}

\bibitem[\protect\citeauthoryear{{Jassal}, {Bagla}  \& {Padmanabhan}}{{Jassal}
  et~al.}{2005}]{2005MNRAS.356L..11J}
{Jassal} H.~K.,  {Bagla} J.~S.,   {Padmanabhan} T.,  2005, \mn@doi [\mnras]
  {10.1111/j.1745-3933.2005.08577.x}, \href
  {https://ui.adsabs.harvard.edu/abs/2005MNRAS.356L..11J} {356, L11}

\bibitem[\protect\citeauthoryear{{Jeffreys}}{{Jeffreys}}{1939}]{jeffreys1998theory}
{Jeffreys} H.,  1939, {Theory of Probability, Oxford University Press, Oxford}

\bibitem[\protect\citeauthoryear{{Kazin} et~al.,}{{Kazin}
  et~al.}{2014}]{Kazin:2014qga}
{Kazin} E.~A.,  et~al., 2014, \mn@doi [\mnras] {10.1093/mnras/stu778}, \href
  {https://ui.adsabs.harvard.edu/abs/2014MNRAS.441.3524K} {441, 3524}

\bibitem[\protect\citeauthoryear{{Kunz}}{{Kunz}}{2012}]{Kunz:2012aw}
{Kunz} M.,  2012, \mn@doi [Comptes Rendus Physique]
  {10.1016/j.crhy.2012.04.007}, \href
  {https://ui.adsabs.harvard.edu/abs/2012CRPhy..13..539K} {13, 539}

\bibitem[\protect\citeauthoryear{{Lewis} \& {Bridle}}{{Lewis} \&
  {Bridle}}{2002}]{Lewis:2002ah}
{Lewis} A.,  {Bridle} S.,  2002, \mn@doi [\prd] {10.1103/PhysRevD.66.103511},
  \href {https://ui.adsabs.harvard.edu/abs/2002PhRvD..66j3511L} {66, 103511}

\bibitem[\protect\citeauthoryear{{Lewis}, {Challinor}  \& {Lasenby}}{{Lewis}
  et~al.}{2000}]{Lewis:1999bs}
{Lewis} A.,  {Challinor} A.,   {Lasenby} A.,  2000, \mn@doi [\apj]
  {10.1086/309179}, \href
  {https://ui.adsabs.harvard.edu/abs/2000ApJ...538..473L} {538, 473}

\bibitem[\protect\citeauthoryear{{Linder}}{{Linder}}{2003}]{Linder:2002et}
{Linder} E.~V.,  2003, \mn@doi [\prl] {10.1103/PhysRevLett.90.091301}, \href
  {https://ui.adsabs.harvard.edu/abs/2003PhRvL..90i1301L} {90, 091301}

\bibitem[\protect\citeauthoryear{{Linder}}{{Linder}}{2005}]{2005PhRvD..72d3529L}
{Linder} E.~V.,  2005, \mn@doi [\prd] {10.1103/PhysRevD.72.043529}, \href
  {https://ui.adsabs.harvard.edu/abs/2005PhRvD..72d3529L} {72, 043529}

\bibitem[\protect\citeauthoryear{{Linder}}{{Linder}}{2006}]{Linder2006OnOD}
{Linder} E.~V.,  2006, \mn@doi [Astroparticle Physics]
  {10.1016/j.astropartphys.2005.12.003}, \href
  {https://ui.adsabs.harvard.edu/abs/2006APh....25..167L} {25, 167}

\bibitem[\protect\citeauthoryear{{Maor}, {Brustein}  \& {Steinhardt}}{{Maor}
  et~al.}{2001}]{2001PhRvL..86....6M}
{Maor} I.,  {Brustein} R.,   {Steinhardt} P.~J.,  2001, \mn@doi [\prl]
  {10.1103/PhysRevLett.86.6}, \href
  {https://ui.adsabs.harvard.edu/abs/2001PhRvL..86....6M} {86, 6}

\bibitem[\protect\citeauthoryear{{Marshall}, {Rajguru}  \& {Slosar}}{{Marshall}
  et~al.}{2006}]{2006PhRvD..73f7302M}
{Marshall} P.,  {Rajguru} N.,   {Slosar} A.,  2006, \mn@doi [\prd]
  {10.1103/PhysRevD.73.067302}, \href
  {https://ui.adsabs.harvard.edu/abs/2006PhRvD..73f7302M} {73, 067302}

\bibitem[\protect\citeauthoryear{{Mehta}, {Cuesta}, {Xu}, {Eisenstein}  \&
  {Padmanabhan}}{{Mehta} et~al.}{2012}]{mehta}
{Mehta} K.~T.,  {Cuesta} A.~J.,  {Xu} X.,  {Eisenstein} D.~J.,   {Padmanabhan}
  N.,  2012, \mn@doi [\mnras] {10.1111/j.1365-2966.2012.21112.x}, \href
  {https://ui.adsabs.harvard.edu/abs/2012MNRAS.427.2168M} {427, 2168}

\bibitem[\protect\citeauthoryear{{Nojiri} \& {Odintsov}}{{Nojiri} \&
  {Odintsov}}{2006}]{Nojiri:2006ww}
{Nojiri} S.,  {Odintsov} S.~D.,  2006, \mn@doi [Physics Letters B]
  {10.1016/j.physletb.2006.04.026}, \href
  {https://ui.adsabs.harvard.edu/abs/2006PhLB..637..139N} {637, 139}

\bibitem[\protect\citeauthoryear{{Pan}, {Saridakis}  \& {Yang}}{{Pan}
  et~al.}{2018}]{2018PhRvD..98f3510P}
{Pan} S.,  {Saridakis} E.~N.,   {Yang} W.,  2018, \mn@doi [\prd]
  {10.1103/PhysRevD.98.063510}, \href
  {https://ui.adsabs.harvard.edu/abs/2018PhRvD..98f3510P} {98, 063510}

\bibitem[\protect\citeauthoryear{{Percival} \& {White}}{{Percival} \&
  {White}}{2009}]{Percival:2008sh}
{Percival} W.~J.,  {White} M.,  2009, \mn@doi [\mnras]
  {10.1111/j.1365-2966.2008.14211.x}, \href
  {https://ui.adsabs.harvard.edu/abs/2009MNRAS.393..297P} {393, 297}

\bibitem[\protect\citeauthoryear{{Perlmutter} et~al.,}{{Perlmutter}
  et~al.}{1998}]{perlmutter1998discovery}
{Perlmutter} S.,  et~al., 1998, \mn@doi [\nat] {10.1038/34124}, \href
  {https://ui.adsabs.harvard.edu/abs/1998Natur.391...51P} {391, 51}

\bibitem[\protect\citeauthoryear{{Planck Collaboration V}}{{Planck
  Collaboration V}}{2020}]{Planck:2019nip}
{Planck Collaboration V} 2020, \mn@doi [\aap] {10.1051/0004-6361/201936386},
  \href {https://ui.adsabs.harvard.edu/abs/2020A&A...641A...5P} {641, A5}

\bibitem[\protect\citeauthoryear{{Planck Collaboration VI}}{{Planck
  Collaboration VI}}{2020}]{Planck:2018vyg}
{Planck Collaboration VI} 2020, \mn@doi [\aap] {10.1051/0004-6361/201833910},
  \href {https://ui.adsabs.harvard.edu/abs/2020A&A...641A...6P} {641, A6}

\bibitem[\protect\citeauthoryear{{Planck Collaboration XIV}}{{Planck
  Collaboration XIV}}{2016}]{Planck:2015bue}
{Planck Collaboration XIV} 2016, \mn@doi [\aap] {10.1051/0004-6361/201525814},
  \href {https://ui.adsabs.harvard.edu/abs/2016A&A...594A..14P} {594, A14}

\bibitem[\protect\citeauthoryear{{Ratra} \& {Peebles}}{{Ratra} \&
  {Peebles}}{1988}]{Ratra:1987rm}
{Ratra} B.,  {Peebles} P.~J.~E.,  1988, \mn@doi [\prd]
  {10.1103/PhysRevD.37.3406}, \href
  {https://ui.adsabs.harvard.edu/abs/1988PhRvD..37.3406R} {37, 3406}

\bibitem[\protect\citeauthoryear{{Rezaei}, {Malekjani}, {Basilakos}, {Mehrabi}
  \& {Mota}}{{Rezaei} et~al.}{2017}]{Rezaei:2017yyj}
{Rezaei} M.,  {Malekjani} M.,  {Basilakos} S.,  {Mehrabi} A.,   {Mota} D.~F.,
  2017, \mn@doi [\apj] {10.3847/1538-4357/aa7898}, \href
  {https://ui.adsabs.harvard.edu/abs/2017ApJ...843...65R} {843, 65}

\bibitem[\protect\citeauthoryear{{Riess} et~al.,}{{Riess}
  et~al.}{1998}]{riess1998observational}
{Riess} A.~G.,  et~al., 1998, \mn@doi [\aj] {10.1086/300499}, \href
  {https://ui.adsabs.harvard.edu/abs/1998AJ....116.1009R} {116, 1009}

\bibitem[\protect\citeauthoryear{{Riess} et~al.,}{{Riess}
  et~al.}{2004}]{2004ApJ...607..665R}
{Riess} A.~G.,  et~al., 2004, \mn@doi [\apj] {10.1086/383612}, \href
  {https://ui.adsabs.harvard.edu/abs/2004ApJ...607..665R} {607, 665}

\bibitem[\protect\citeauthoryear{{Ross}, {Samushia}, {Howlett}, {Percival},
  {Burden}  \& {Manera}}{{Ross} et~al.}{2015}]{Ross:2014qpa}
{Ross} A.~J.,  {Samushia} L.,  {Howlett} C.,  {Percival} W.~J.,  {Burden} A.,
  {Manera} M.,  2015, \mn@doi [\mnras] {10.1093/mnras/stv154}, \href
  {https://ui.adsabs.harvard.edu/abs/2015MNRAS.449..835R} {449, 835}

\bibitem[\protect\citeauthoryear{{Scolnic} et~al.,}{{Scolnic}
  et~al.}{2018}]{2018ApJ...859..101S}
{Scolnic} D.~M.,  et~al., 2018, \mn@doi [\apj] {10.3847/1538-4357/aab9bb},
  \href {https://ui.adsabs.harvard.edu/abs/2018ApJ...859..101S} {859, 101}

\bibitem[\protect\citeauthoryear{{Seljak} et~al.,}{{Seljak}
  et~al.}{2005}]{SDSS:2004kqt}
{Seljak} U.,  et~al., 2005, \mn@doi [\prd] {10.1103/PhysRevD.71.103515}, \href
  {https://ui.adsabs.harvard.edu/abs/2005PhRvD..71j3515S} {71, 103515}

\bibitem[\protect\citeauthoryear{{Staicova} \& {Benisty}}{{Staicova} \&
  {Benisty}}{2022}]{2022A&A...668A.135S}
{Staicova} D.,  {Benisty} D.,  2022, \mn@doi [\aap]
  {10.1051/0004-6361/202244366}, \href
  {https://ui.adsabs.harvard.edu/abs/2022A&A...668A.135S} {668, A135}

\bibitem[\protect\citeauthoryear{{Trotta}}{{Trotta}}{2007}]{2007MNRAS.378...72T}
{Trotta} R.,  2007, \mn@doi [\mnras] {10.1111/j.1365-2966.2007.11738.x}, \href
  {https://ui.adsabs.harvard.edu/abs/2007MNRAS.378...72T} {378, 72}

\bibitem[\protect\citeauthoryear{{Vikman}}{{Vikman}}{2005}]{2005PhRvD..71b3515V}
{Vikman} A.,  2005, \mn@doi [\prd] {10.1103/PhysRevD.71.023515}, \href
  {https://ui.adsabs.harvard.edu/abs/2005PhRvD..71b3515V} {71, 023515}

\bibitem[\protect\citeauthoryear{{Weinberg}}{{Weinberg}}{1989}]{Weinberg:1988cp}
{Weinberg} S.,  1989, \mn@doi [Reviews of Modern Physics]
  {10.1103/RevModPhys.61.1}, \href
  {https://ui.adsabs.harvard.edu/abs/1989RvMP...61....1W} {61, 1}

\bibitem[\protect\citeauthoryear{{Wetterich}}{{Wetterich}}{1988}]{Wetterich:1987fm}
{Wetterich} C.,  1988, \mn@doi [Nuclear Physics B]
  {10.1016/0550-3213(88)90193-9}, \href
  {https://ui.adsabs.harvard.edu/abs/1988NuPhB.302..668W} {302, 668}

\bibitem[\protect\citeauthoryear{{Xia}, {Zhao}, {Li}, {Feng}  \& {Zhang}}{{Xia}
  et~al.}{2006}]{2006PhRvD..74h3521X}
{Xia} J.-Q.,  {Zhao} G.-B.,  {Li} H.,  {Feng} B.,   {Zhang} X.,  2006, \mn@doi
  [\prd] {10.1103/PhysRevD.74.083521}, \href
  {https://ui.adsabs.harvard.edu/abs/2006PhRvD..74h3521X} {74, 083521}

\bibitem[\protect\citeauthoryear{{Zuntz} et~al.,}{{Zuntz}
  et~al.}{2015}]{Zuntz:2015dhn}
{Zuntz} J.,  et~al., 2015, \mn@doi [Astronomy and Computing]
  {10.1016/j.ascom.2015.05.005}, \href
  {https://ui.adsabs.harvard.edu/abs/2015A&C....12...45Z} {12, 45}

\makeatother
\end{thebibliography}
\label{lastpage}
\end{document}